\documentclass[]{spie}  

\usepackage{amsmath,amsfonts,amssymb}
\usepackage{graphicx}
\usepackage[colorlinks=true, allcolors=blue]{hyperref}

\usepackage{upgreek} 
\usepackage{booktabs}
\usepackage[caption=false]{subfig}

\newcommand{\ka}{K$\upalpha$ }
\newcommand{\kb}{K$\upbeta$ }

\graphicspath{{./}{Figures/}}

\title{Measuring the Quantum Efficiency of X-Ray Hybrid CMOS Detectors}

\author[$*$a]{Joseph M. Colosimo}
\author[a]{Abraham D. Falcone}
\author[a]{Mitchell Wages}
\author[a,b]{Samuel V. Hull}
\author[a]{David N. Burrows}
\author[a]{Mitchell Range}
\author[a]{Fredric Hancock}
\author[a]{Cole R. Armstrong}
\author[a]{Gooderham McCormick}
\author[a]{Daniel M. LaRocca}
\affil[a]{The Pennsylvania State University, 525 Davey Lab, University Park, Pennsylvania, United States}
\affil[b]{Goddard Space Flight Center, 8800 Greenbelt Rd, Greenbelt, Maryland, United States}

\authorinfo{*jcolosimo@psu.edu}

\begin{document} 
\maketitle

\begin{abstract}
Next-generation X-ray observatories, such as the Lynx X-ray Observatory Mission Concept, will require detectors with high quantum efficiency (QE) across the soft X-ray band to observe the faint objects that drive their mission science cases.
Hybrid CMOS Detectors (HCDs), a form of active-pixel sensor, are promising candidates for use on these missions because of their faster read-out, lower power consumption, and greater radiation hardness than detectors used in the current generation of X-ray telescopes. 
In this work, we present QE measurements of a Teledyne H2RG HCD.
These measurements were performed using a gas-flow proportional counter as a reference detector to measure the absolute flux incident on the HCD. 
We find an effective QE of $95.0 \pm 1.1\%$ at the Mn K$\upalpha$/K$\upbeta$ lines (at 5.9 and 6.5 keV), $98.5 \pm 1.8\%$ at the Al \ka line (1.5 keV), and $85.0 \pm 2.8\%$ at the O \ka line (0.52 keV). 
\end{abstract}

\keywords{CMOS sensors, Hybrid CMOS, Soft X-ray, X-ray detectors,  Quantum efficiency}

\section{INTRODUCTION}
\label{sec:intro}  

The Lynx X-Ray Observatory Mission Concept  \cite{Lynx19, Gaskin19}, along with other proposed and planned next-generation X-ray observatories, will provide unprecedented observations of the early universe.
These missions will require large effective areas to enable detections of distant, faint sources. 
Detectors with capabilities surpassing those of the charge-coupled devices (CCDs) used in the focal planes of current missions will be required for future telescopes.
These detectors will need high quantum efficiency (QE), the fraction of incident X-rays individually detected, in order to fully utilize the large collecting areas provided by the mirror assemblies and maximize sensitivity to faint sources.
Future missions also require detectors with fast read-out in order to avoid pile-up, an effect in which multiple X-rays land in the same pixel during a single frame, when observing bright sources.
Without a major increase in the detector frame rate, pile-up would be exacerbated by the greater effective areas of these missions. 

\begin{figure}[bt]
    \centering
    \includegraphics[width=0.4\textwidth]{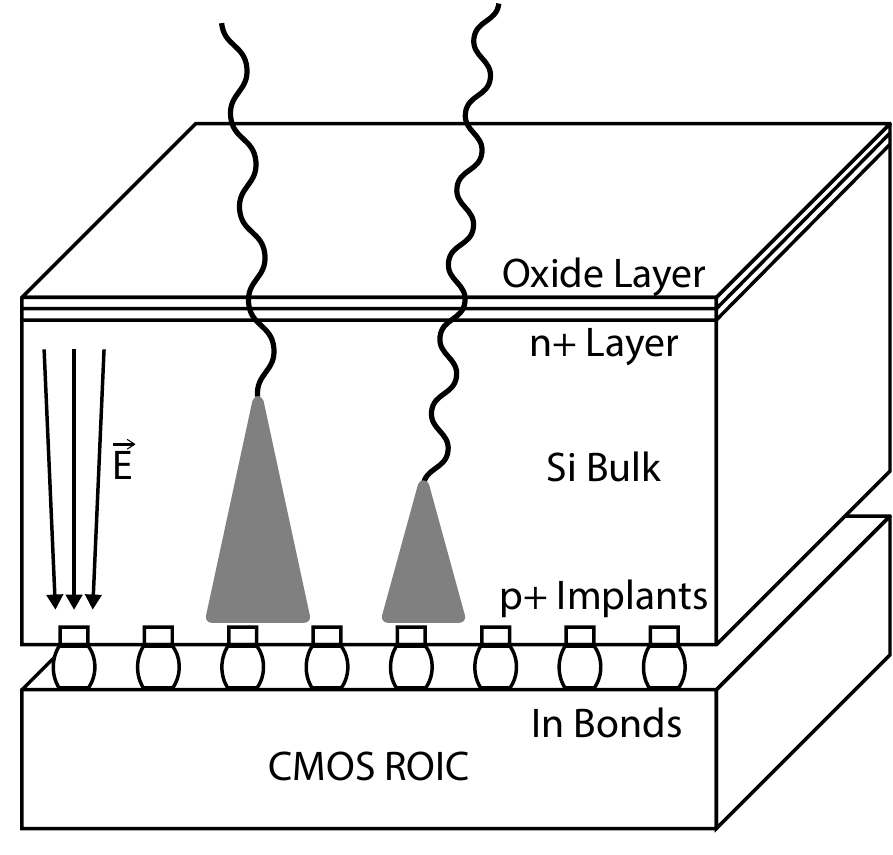}
    \caption{Diagram showing the structure of silicon HCDs. In these detectors, X-rays  absorbed in the absorber layer create a charge cloud which is collected and transferred to the CMOS ROIC to be measured and read out. Adapted from Ref.~\citenum{Bongiorno15}.}
    \label{fig:detector_diagram}
\end{figure}

Hybrid CMOS Detectors (HCDs) are promising candidates for use in these future X-ray observatories.
HCDs are composed of a silicon absorber layer bonded to a complementary metal–oxide–semiconductor (CMOS) read-out integrated circuit (ROIC). The absorber layer produces photoelectrons when absorbing an X-ray and the ROIC reads the amount of charge produced. 
Each pixel in an HCD contains its own ROIC, removing the need to transfer charge through other pixels and providing significantly faster readout, intrinsic radiation hardness, and lower power demands compared to the CCDs on current missions.
This hybridized structure, shown in Fig.~\ref{fig:detector_diagram}, allows for independent optimization of the ROIC and absorber layers during fabrication.
HCDs provide high QE across the soft X-ray band, due to their thin inactive surface layers and large depletion depths ($100\;\upmu$m in current models) enabled by the use of high-resistivity silicon in the absorber layer. 

While the Penn State High-Energy Astrophysics Detector and Instrumentation Laboratory has previously measured a high QE on an H1RG HCD at the Mn K$\upalpha$/K$\upbeta$ lines \cite{Bongiorno15}, further measurements are necessary to characterize the QE on other detectors and over a range of energies.
Characterization of the QE at low energies (0.2 - 1.5 keV) is crucial to detector development, as good low-energy performance is vital to the science goals of future missions. 
For example, high QE at low energies would allow Lynx to more efficiently observe high-redshift quasars, whose power-law spectra have significant contribution in this band, and allow imaging of the halos of nearby galaxies, with low-energy thermal emission peaks \cite{Lynx19}.
In this work, we present measurements of the QE of an H2RG HCD at moderate and low energies. 
In particular, we report on the QE of this detector at the Mn K$\upalpha$/K$\upbeta$ (5.90/6.49 keV), Al \ka (1.49 keV), and O \ka (0.52 keV) characteristic X-ray emission lines. 
We also present plans for a QE measurement at the C \ka line (0.28 keV), and for additional QE measurements for the more recent generation of X-ray HCDs.

\section{QUANTUM EFFICIENCY MODEL}
\label{sec:model}

\begin{figure}[bt]
    \centering
    \includegraphics[width=0.65\textwidth]{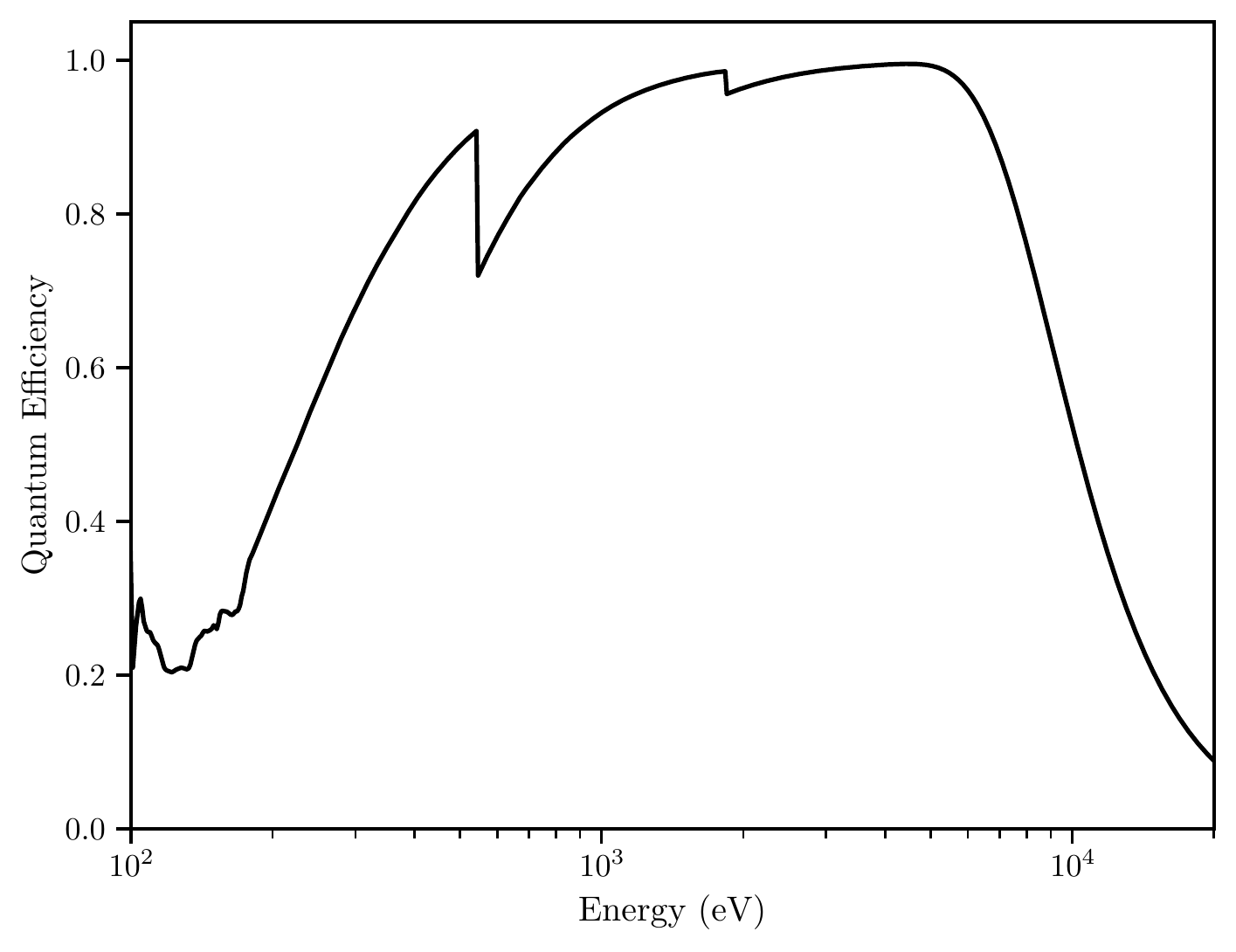}
    \caption{Estimate of the H2RG QE based on the 1D slab absorption model.}
    \label{fig:qe_model}
\end{figure}

We use a 1D slab absorption model to estimate the QE of HCDs. 
In this model, the QE is given by the fraction of X-rays transmitted through inactive surface layers and absorbed by the silicon absorber layer. 
This can be expressed in the following form:
\begin{equation}
    \textrm{QE}\,(E) = \left(1-e^{-\sigma_{\textrm{\tiny{Si}}}(E)\, n_{\textrm{\tiny{Si}}} t_{\textrm{\tiny{abs}}}}\right)\prod^{N}_{i}e^{-\sigma_{i}(E)\, n_{i} t_{i}}
    \label{eq:model}
\end{equation}
where $t_\textrm{abs}$ is the thickness of the absorbing layer (with the cross section and number density for silicon) and the product is over the inactive surface layers with atomic cross sections $\sigma_i$, number densities $n_i$, and thicknesses $t_i$. 
The H2RG has a 100 $\upmu \rm{m}$ Si absorber layer and a single inactive surface layer composed of 0.1 $\upmu \rm{m}$ SiO$_2$. In this work, we calculate transmissions using values provided by Ref.~\citenum{Henke93}. The resulting H2RG QE model is shown in Fig.~\ref{fig:qe_model}.

At energies below 5 keV, few X-rays are transmitted through the Si absorbing layer, and the QE is primarily determined by the transmittance of the inactive surface layer. 
Above 5 keV, almost all X-rays are transmitted through the inactive surface layer, and the QE is primarily determined by the fraction of X-rays absorbed by the Si absorbing layer.  
While the simple slab absorption model has proved accurate in past HCD QE measurements \cite{Bongiorno15, Prieskorn14, Kenter05}, effects ignored by this simple 1D model (e.g., charge diffusion or pixel-to-pixel variations) may cause deviations from model values. 
In addition, uncertainties in surface layer thicknesses could cause the low-energy QE to deviate from the modeled values.
QE measurements are therefore necessary to characterize the response of an X-ray detector over a range of energies, to determine their suitability for future missions, and to guide the development of future detectors.

\section{EXPERIMENTAL METHODS}
\label{sect:methods}

QE measurements require a separate, well-characterized, reference detector to determine the incident X-ray flux.
We used a gas-flow proportional counter (PC) as our reference detector because of its known QE and high response across the soft X-ray band. 
During the experiment, both the H2RG and PC were simultaneously exposed at a similar distance to the same source.
The flux measured by the H2RG was compared to the flux measured by the reference detector to determine the QE.
The QE measurements were conducted in a 47-meter X-ray beamline operated by the Penn State High-Energy Astrophysics Detector and Instrumentation Laboratory.
The mounting structure containing both detectors can be seen in and outside of the beamline in Fig.~\ref{fig:detectors}.
The large distance between the source and detectors as well as the simultaneous data collection minimized the difference in flux incident upon both detectors.
The beamline is maintained at high vacuum ($<10^{-5}$ torr) to minimize attenuation of X-rays and avoid the condensation of water vapor onto the HCD (which is cooled during operation).

\subsection{H2RG}
\label{sec:h2rg}

We conducted our QE measurements on a modified X-ray H2RG fabricated by Teledyne Imaging Sensors.
This detector uses a standard H2RG ROIC, with $2048\times2048$ 18 $\upmu$m pitch pixels, bonded to a $1024\times1024$ array of 36 $\upmu$m pitch Si absorber cells, with only out of every four ROIC pixels connected to the absorber layer. 
This design was selected to increase the distance between pixels in order to reduce the effect of interpixel capacitance, a form of crosstalk which decreased the energy resolution of older-generation H1RG HCDs \cite{Falcone12}.   
While extensive characterization measurements have been taken for this detector \cite{Prieskorn13}, and it even has flight heritage on the Water Recovery X-ray Rocket mission \cite{Wages19}, no QE measurements had been made on the X-ray H2RG prior to this work.

During QE measurements, the detector was cooled to 160 K using a copper strap thermally connecting the device to an internal liquid nitrogen dewar. 
This temperature is slightly higher than the optimal operating temperature of $\lesssim 150$ K, leading to increased noise and reduced energy resolution; however, we found that operating below 160 K allowed for deposition of an ice layer on the detector surface, which reduced the measured QE. 
This effect was observed as a gradual degradation of the QE with time.
We installed a residual gas analyzer to measure the partial pressure of water vapor in the chamber and confirmed that it was lower than the deposition pressure at 160 K during our QE measurements. 

The H2RG was operated in ``up-the-ramp'' mode, in which many frames were taken successively without clearing charge. The charge accumulated in a single frame was found by taking the difference in pixel values between two consecutive frames. Row and channel noise were then removed from the image using a boxcar smoothing algorithm.
In order to identify X-ray events, we found pixels which had charge depositions that were local maxima and above a set primary threshold. 
If charge in a surrounding pixel exceeded a secondary threshold, it was included in the total charge of the X-ray event.

This H2RG is an engineering-grade detector. 
As such, it has many charge traps which create areas with significant dark current and limited dynamic range.
We chose to exclude X-ray events from pixels in regions associated with these charge traps, as these pixels may have reduced sensitivity to incident X-rays.
A science-grade detector would have fewer defects, so this measurement better reflects the expected QE of the detectors which would be used on a mission.
We also excluded events from pixels too close to the edge of the detector, where edge effects reduce the detector sensitivity. 
After masking off these pixels, we are left with 85.7\% of the total detector area as active. 
The area of the detector was scaled by this value in calculations of the measured flux and the QE.

\subsection{Proportional Counter}
\label{sec:pc}

\begin{figure}[bt]
    \centering
    \subfloat[]{%
      \includegraphics[width=0.45\textwidth]{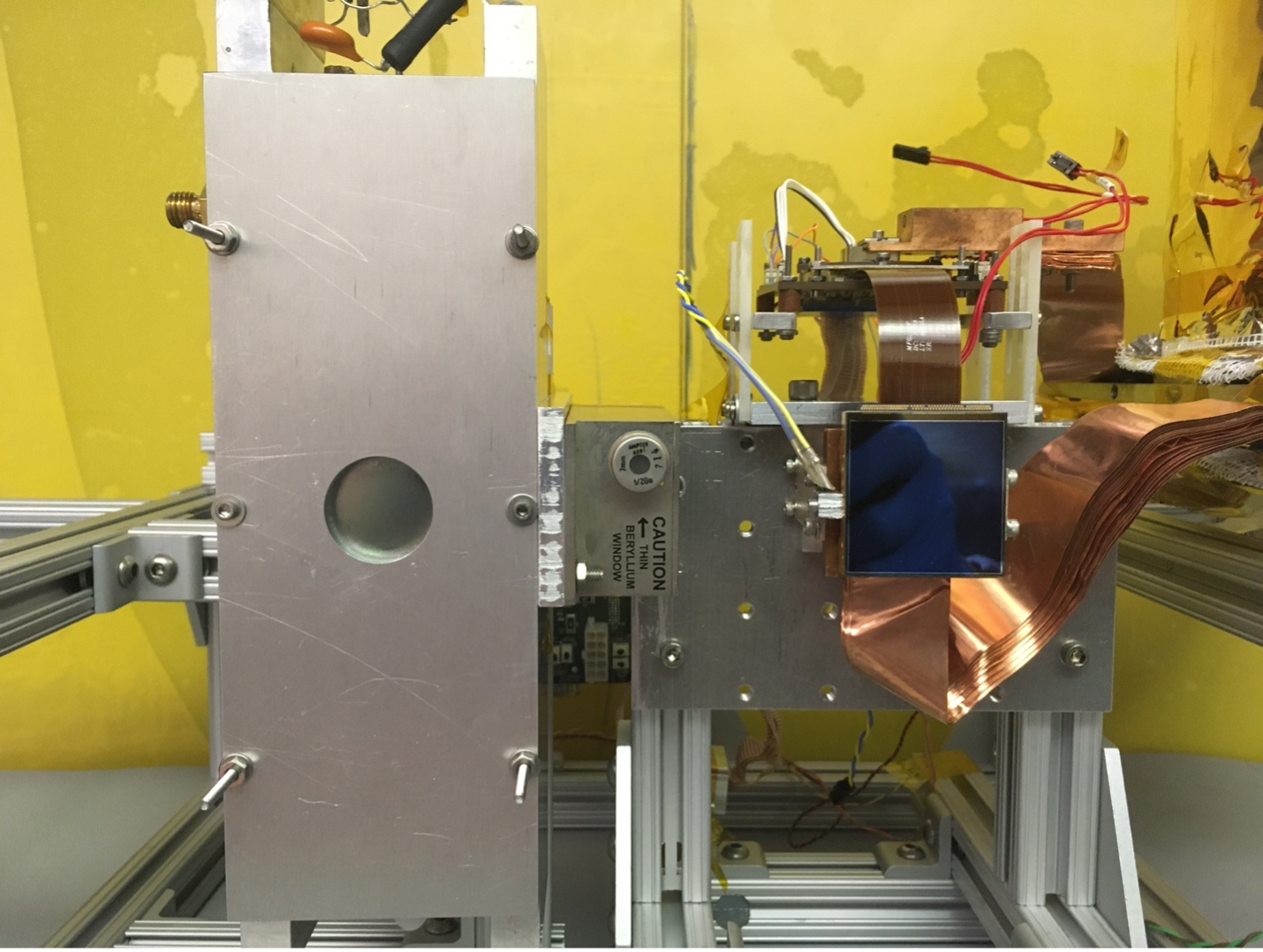}%
    }\quad
    \subfloat[]{%
      \includegraphics[width=0.45\textwidth]{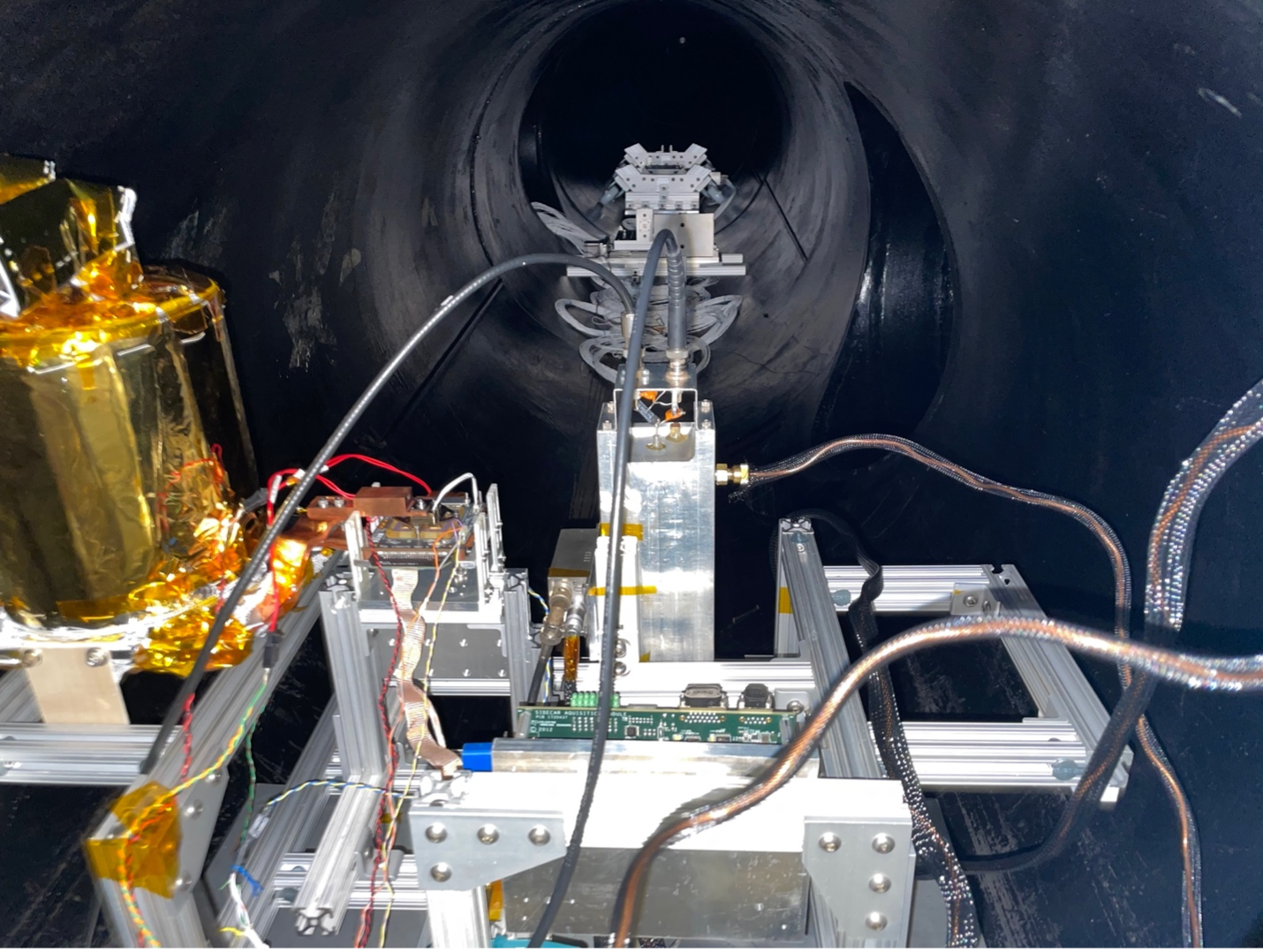}
    }
    \caption{The front of the detector mounting structure shown outside the vacuum chamber (a), with the PC (left) and the H2RG (right), and the back of the same structure shown inside of the beamline where the QE measurements took place (b). A Si PIN diode detector was also mounted between these detectors but was not used during QE measurements.}
    \label{fig:detectors}
\end{figure}

The PC and H2RG were mounted side-by-side and equidistant from the midline of the chamber, as shown in Fig.~\ref{fig:detectors}. 
This arrangement provides roughly equal flux incident on both detectors. 
P-10 gas (
90\% argon and 10\% methane) was flowed through the PC at a rate of $\sim200$ cc/min.
The temperature and pressure of the gas were measured to ensure accurate values for calculation of the PC QE.
The gas is separated from vacuum by a thin window installed in the aperture of the detector. 
The window was manufactured by Luxel for high transmission of low-energy X-rays and was constructed from aluminized polyimide, supported by a stainless steel mesh.

In order for the PC to act as a reference detector and accurately measure the absolute flux produced by the source, its QE must be well-characterized.
PCs are commonly used as reference detectors because their QE can be reliably determined through measurements of gas properties and window transmission.
We calculated the QE using a model similar to the HCD model described in \S\ref{sec:model}, where the PC QE is given by the fraction of X-rays transmitted through the window which are then absorbed by the P-10 gas. 
The gas absorption was calculated based on the pressure, temperature, and composition of the P-10 gas, while the transmissivity of the window in the PC aperture was measured to ensure accuracy. 

Measurements of the window transmittance were necessary to obtain an accurate absolute calibration of the PC, as slight variations in the thickness of the window layers can cause significant changes in its transmittance. 
In the low-energy band ($E\lesssim 1.5$ keV), almost all X-rays transmitted by the window are absorbed by the P-10 gas, so the QE of the detector is primarily a function of the window transmission. 
Calibration measurements are also important at high energies, as the stainless steel mesh used to support the window acts as a binary mask, blocking X-rays at all energies throughout the soft X-ray band.

\subsubsection{Proportional Counter Calibration}
\label{sec:pc_calibration}

We measured the transmittance of the PC window using a secondary window which was also manufactured by Luxel with the same specifications as the primary window. 
The secondary window was installed in the aperture of the PC during the calibration measurements. 
We then mounted the primary window directly in front of the secondary window and compared the count rates when the primary window was present and when it was removed.
The ratio of the observed count rates in these measurements gives the transmittance of the window at a given energy. 

\begin{figure}[bt]
    \centering
    \includegraphics[width=0.65\textwidth]{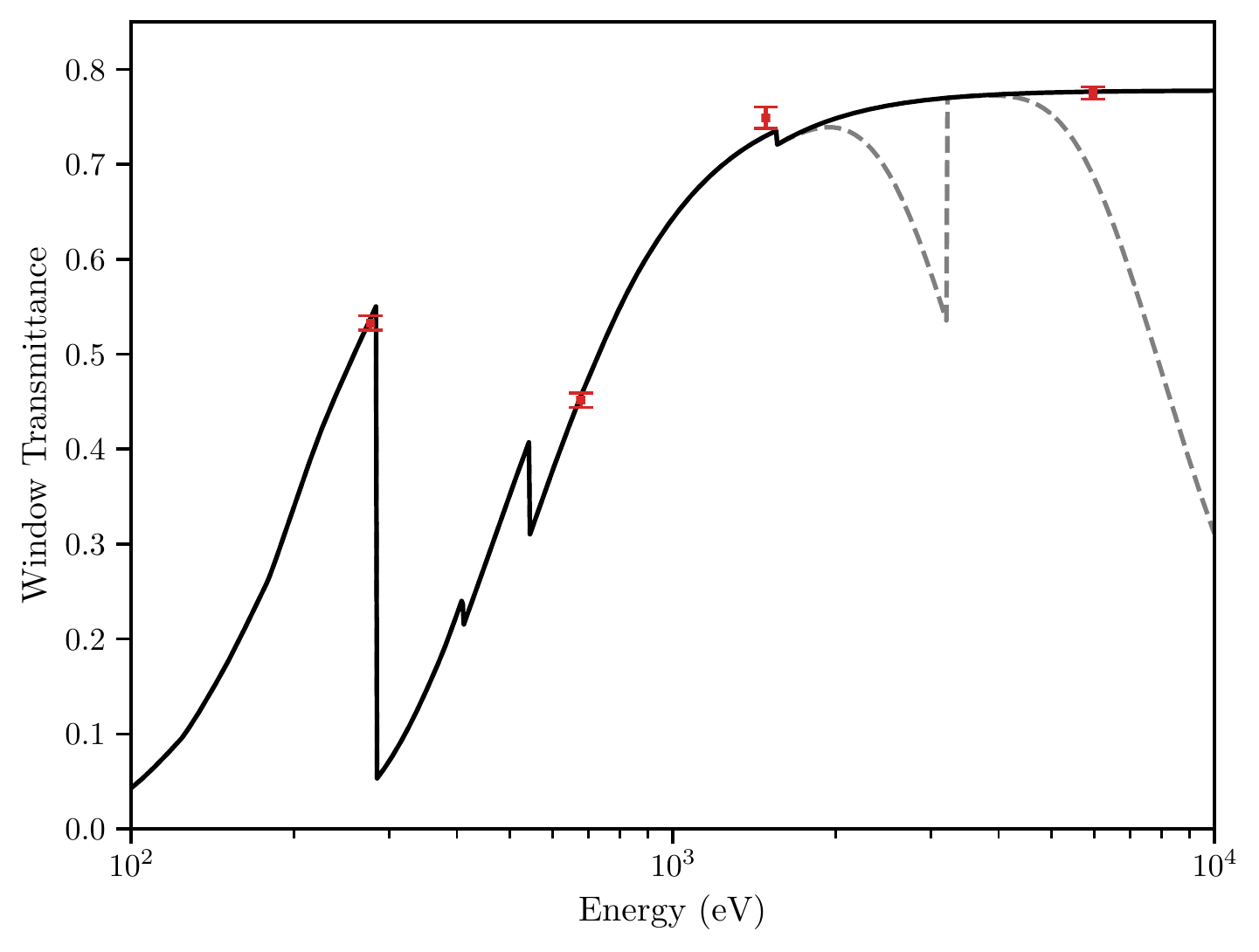}
    \caption{Measurements of PC window transmittance (red markers) shown on top of the fitted transmittance model (black line). The PC QE is shown by the dotted gray line. Note that the QE and window transmittance only diverge at higher energies, where a significant fraction of X-rays are transmitted through the gas.}
    \label{fig:pc_window}
\end{figure}

We used radioactive sources because these provide a known flux, decaying at a known rate.
An Fe-55 source was used for a measurement of the effective window transmission at the Mn K$\upalpha$/K$\upbeta$ energies. We made transmission measurements at lower energies using targets fluoresced by Polonium-210 alpha particle sources. 
A fluoresced aluminum target was used to make transmission measurements at the Al \ka line (1.49 keV). Measurements were also made at the F \ka (0.68 keV) and C \ka (0.28 keV) lines, using Teflon (C$_2$F$_4$) and polyethylene (C$_2$H$_4$) targets. This was done in preparation for future QE measurements at lower energies and to effectively constrain a model of window transmission. 

We fit a model to the measured transmission values in order to obtain the window transmission for all energies. In our model, the window layer thicknesses were fixed at the values provided by Luxel, and only the mesh transmittance was allowed to vary.
We found that this fitted model agrees quite well with our measurements and used it to determine the window transmittance at the O \ka energy. 
This fit is shown in Fig.~\ref{fig:pc_window}, along with the  the measured transmittance values and the modeled PC QE.

\subsection{X-Ray Sources}
\label{sec:sources}

We used two different sources of X-rays for our QE measurements.  A radioactive Fe-55 source produced X-rays at the Mn \ka and \kb energies, and a Manson electron-impact X-ray source produced lower-energy X-rays at the Al and O \ka lines. These sources required careful calibration to ensure that both the energies and spatial distribution of the X-rays were well known. Both sources and their calibrations are described in this section.

\subsubsection{Fe-55 Source}
\label{sec:fe55}

We used an Fe-55 radioactive source to produce emission at 5.9 and 6.5 keV.
We mounted the Fe-55 source on the detector end of the beamline (approximately 2.4 m from the detectors) in order to produce a sufficient count rate. 
Unlike the Manson source, the Fe-55 source emits isotropically, so the incident flux is only a function of distance to the source.
The Fe-55 source is not exactly equidistant from both detectors, leading the flux incident on the HCD to be  $0.18\pm0.03\%$ greater than the flux incident on the PC. 
The source was placed on an electronic linear stage so that it could be moved behind a shutter when not in use, allowing for dark or Manson measurements without opening the chamber. 

The PC does not have sufficient energy resolution to distinguish between the \ka and \kb peaks. Because the QE of the PC differs significantly at these energies (as seen in Fig.~\ref{fig:pc_window}), knowing the fraction of incident X-rays at each energy is essential to accurately determining the fraction of X-rays absorbed by the P-10 gas.
The intrinsic ratio of Mn \kb to \ka emission is 0.1195 \cite{scofield74}; however, the observed line ratio is increased by the greater transmission of \kb X-rays by Fe-55 source and its casing. 
We calculated the ratio of the emitted \kb to \ka X-rays to be $0.132\pm0.002$ based on the thicknesses of the Fe-55 source layer and Ni casing provided by the manufacturer. 

\subsubsection{Manson Source}
\label{sec:manson}

A Manson ultra-soft X-ray source, a type of electron-impact source, was used to produce X-rays for measurements at lower energies.
We used an aluminum anode in the Manson source for QE measurements at the Al \ka line  (1.5 keV) and a heavily oxidized magnesium anode for measurements at the O \ka line (0.52 keV). 
In addition to characteristic lines, the Manson source also emits continuum X-rays. These X-rays must be suppressed in order to prevent X-rays at different energies from being included in the characteristic X-ray fluxes used in the QE measurements. 
We installed carefully-designed filters in the source end of the beamline to block continuum X-rays. 
We used an aluminum filter, constructed from a piece of heavy-duty aluminum foil, for the Al \ka measurements and a filter constructed from thin chromium foil supported by mylar for the O \ka measurements. 
While the filters do not completely eliminate the continuum emission, they provide sufficient separation between the characteristic X-ray and continuum peaks so that both can be modeled and the counts can be separated.
These filters also block optical light produced by the Manson filament, which would be registered by the detector (adding noise to our measurements) and could potentially damage the CMOS ROIC.

While the large distance between the Manson source and the detectors ($\sim47$ m) allowed for fairly uniform flux incident on the detectors, we found that it was neither perfectly uniform nor perfectly symmetric. We obtained a calibration for both Manson source/filter configurations using a detector mounted on a linear stage. We measured the count rate at the position of both the H2RG and PC. These measurements were repeated to account for variability in the absolute flux produced by the Manson source. We found that in the Al \ka configuration, the H2RG received $99.1\pm1.0\%$ of the flux received by the PC. In the O \ka configuration, the H2RG received $100.0\pm0.2\%$ of the flux received by the PC.

\begin{figure}[bt]
    \centering
    \subfloat[H2RG]{%
      \includegraphics[width=0.48\textwidth]{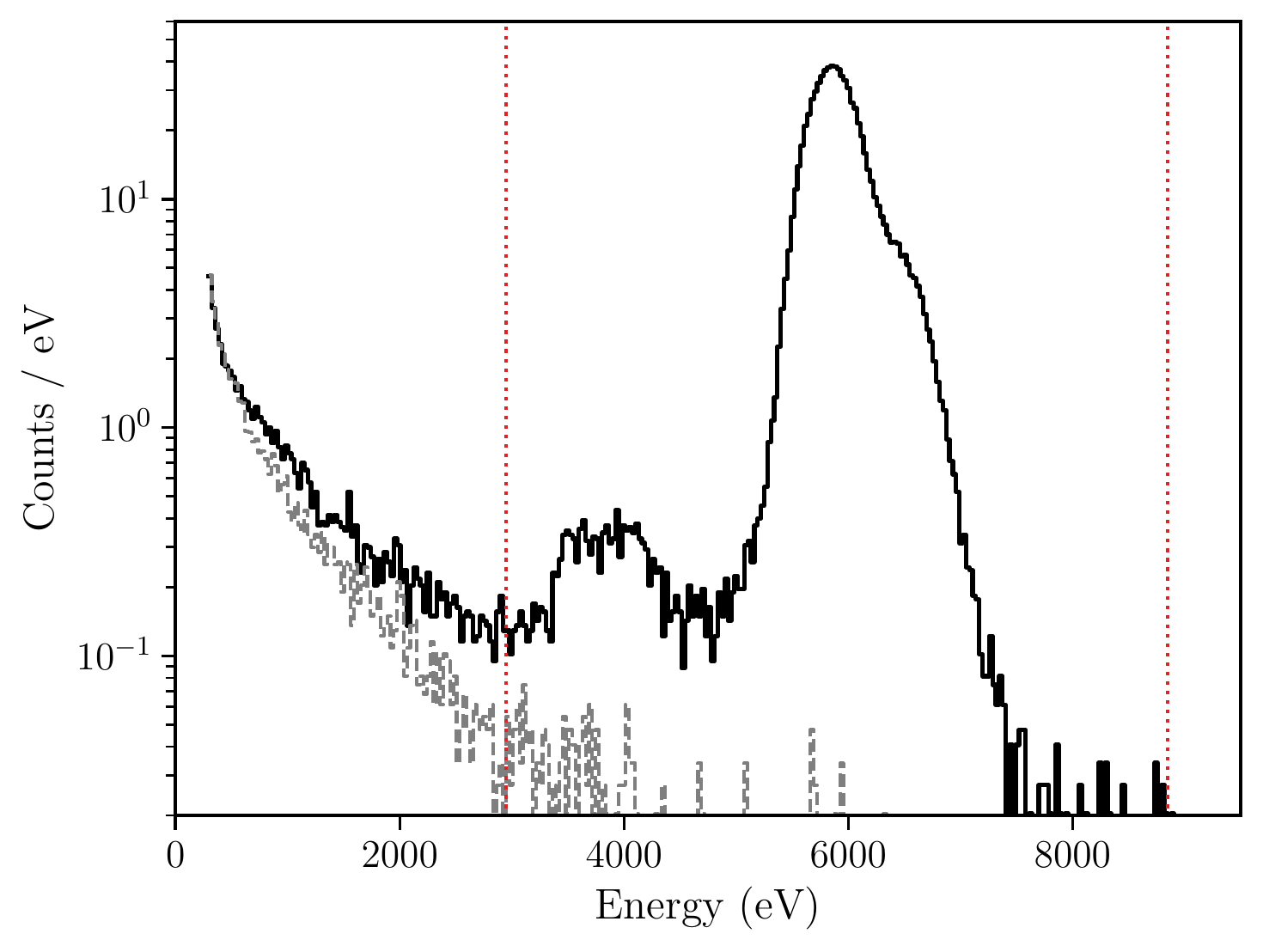}%
    }\quad
    \subfloat[PC]{%
      \includegraphics[width=0.48\textwidth]{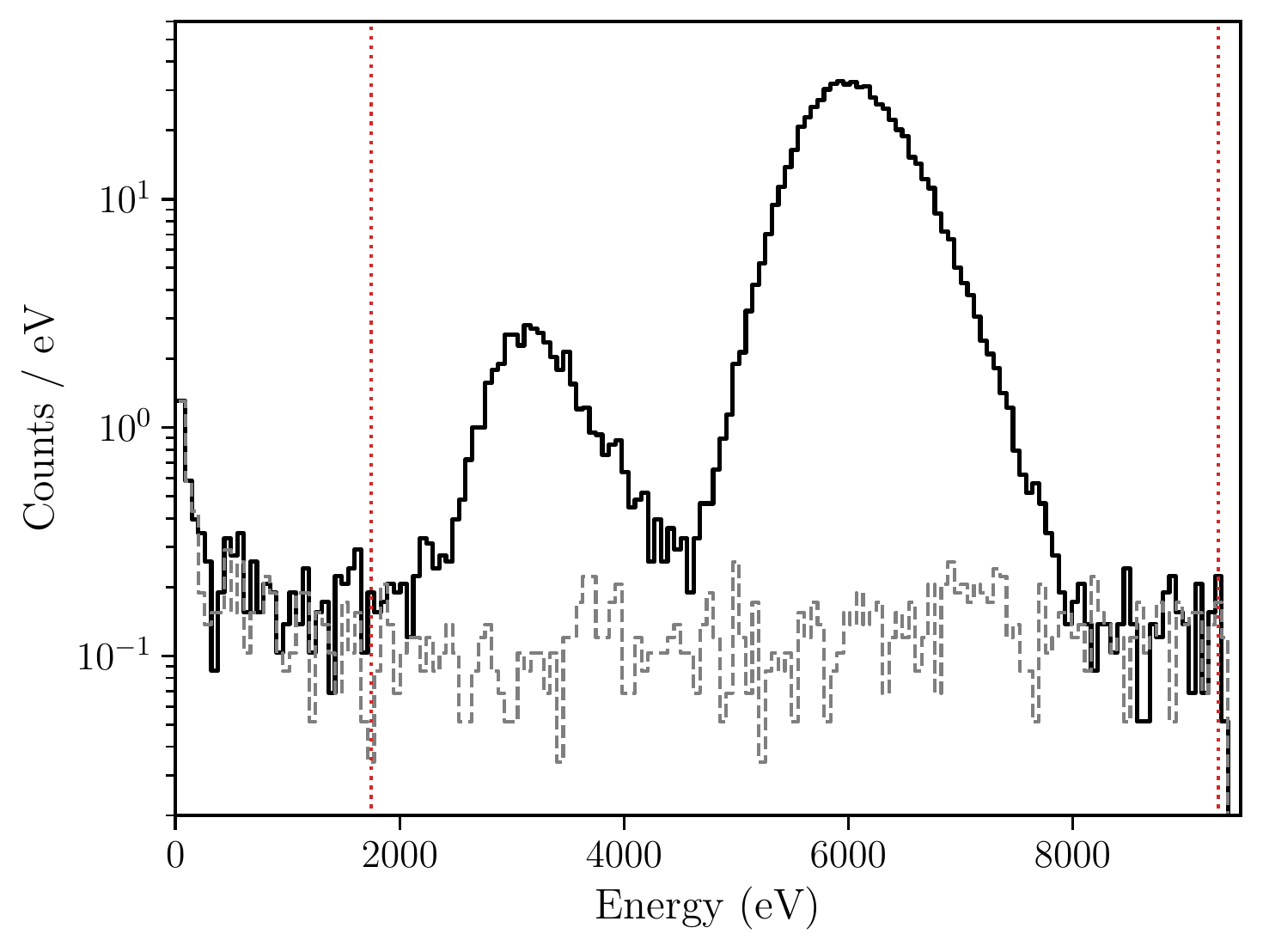}
    }
    \vspace{0.25 cm}
    \caption{H2RG and PC spectra from the Fe-55 QE measurements. Both spectra include combined Mn K$\upalpha$/K$\upbeta$ peaks with escape peaks at lower energies. The (scaled) background spectra from the dark measurement are shown by the gray dashed lines. The bounds used to sum the counts and determine the fluxes are shown by the red dotted lines. In each of the spectra, the energy conversion from H2RG ADU and PC channel are determined by fitting for the peak location.}
    \label{fig:spec_Mn}
\end{figure}

\section{ANALYSIS AND RESULTS}
\label{sec:analysis}

During data acquisition, both detectors measured the incident flux produced by a given source simultaneously. 
At each of the energies measured, we recorded data with the detectors exposed to the source then subsequently recorded background measurements with the source blocked. 
Each measurement lasted for a duration of approximately 10 minutes (with exact live times for the detectors differing slightly). 
In this section, we describe the methods used to analyze this data, discuss the QE calculation, and present the results of our measurements.

\subsection{Mn \texorpdfstring{K$\upalpha$/K$\upbeta$}{Ka/Kb} Measurement}

Fig.~\ref{fig:spec_Mn} shows the H2RG and PC spectra from the Fe-55 QE measurement.
Due to the poor energy resolution of the PC and H2RG (when operated at 160 K), the Mn \ka and \kb peaks are blended together in both spectra. 
For this reason, we make an effective QE measurement at the combined Mn \ka and \kb energies.

We determine the count rates for both detectors by summing all counts between specified bounds, shown in red in Fig.~\ref{fig:spec_Mn}. 
The sums of the counts in the background spectra within these bounds are subtracted from these count rates, removing the instrumental backgrounds and leaving only the count rates from real X-ray events. 
Spectra from both detectors include escape peaks, as the Mn K$\upalpha$/K$\upbeta$ X-rays are more energetic than the binding energies of the Ar and Si K shells. 
The escape peaks were included within the bounds of the sums because they correspond to real Mn K X-rays registered by the detectors.

\subsection{Al \texorpdfstring{\ka}{Ka} and O \texorpdfstring{\ka}{Ka} Measurements}

\begin{figure}[bt]
    \centering
    \subfloat[H2RG]{%
      \includegraphics[width=0.48\textwidth]{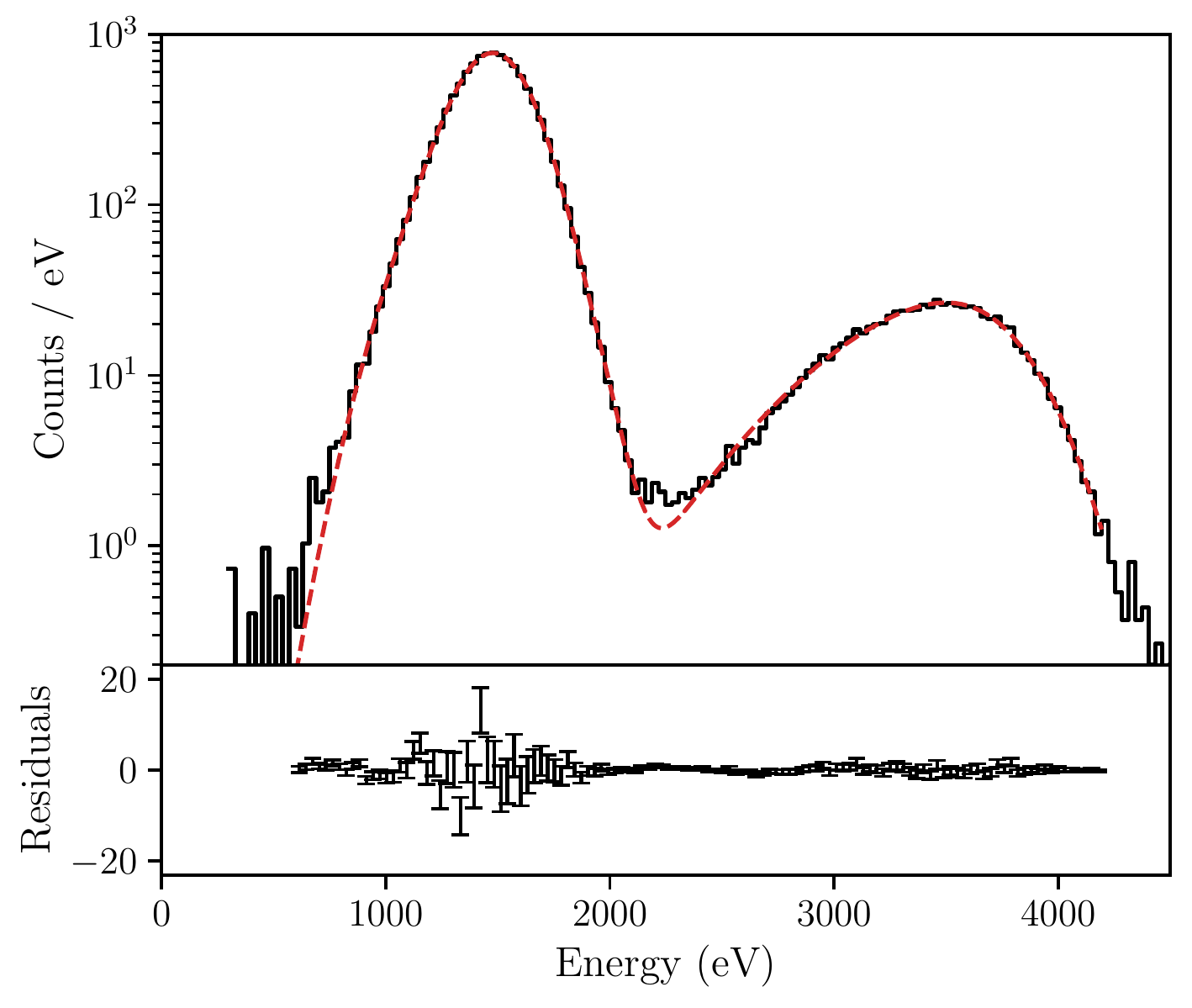}%
    }\quad
    \subfloat[PC]{%
      \includegraphics[width=0.48\textwidth]{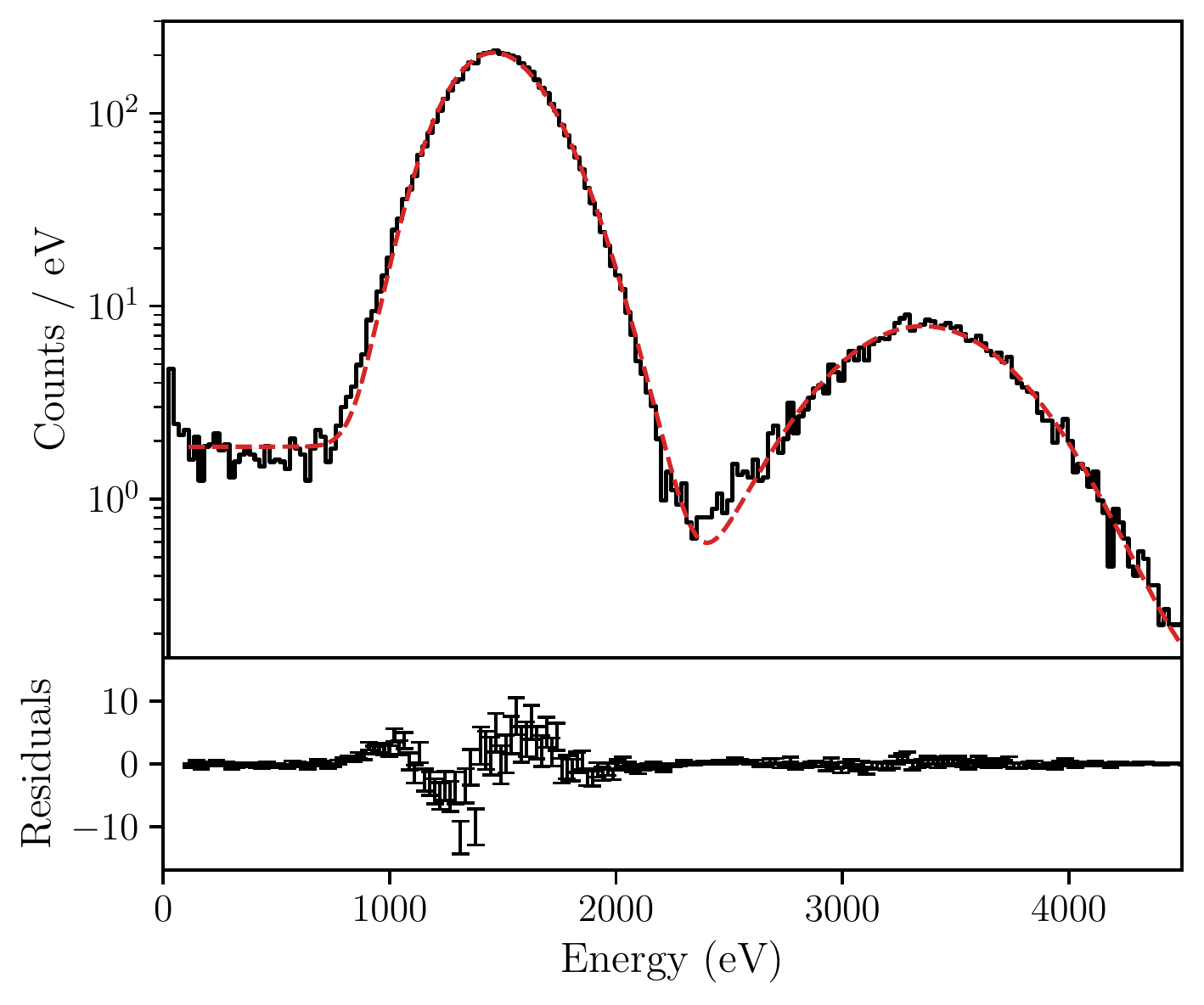}
    }
    \vspace{0.25 cm}
    \caption{H2RG and PC spectra from the Al \ka QE measurements. The H2RG spectrum has been background subtracted in order to remove instrument background counts from this measurement. The fitted models are shown by the red dashed line with residuals shown below. The lower-energy peaks are from the Al \ka counts, while the higher-energy peaks are from continuum emission.}
    \label{fig:spec_Al}
\end{figure}

\begin{figure}[bt]
    \centering
    \subfloat[H2RG]{%
      \includegraphics[width=0.48\textwidth]{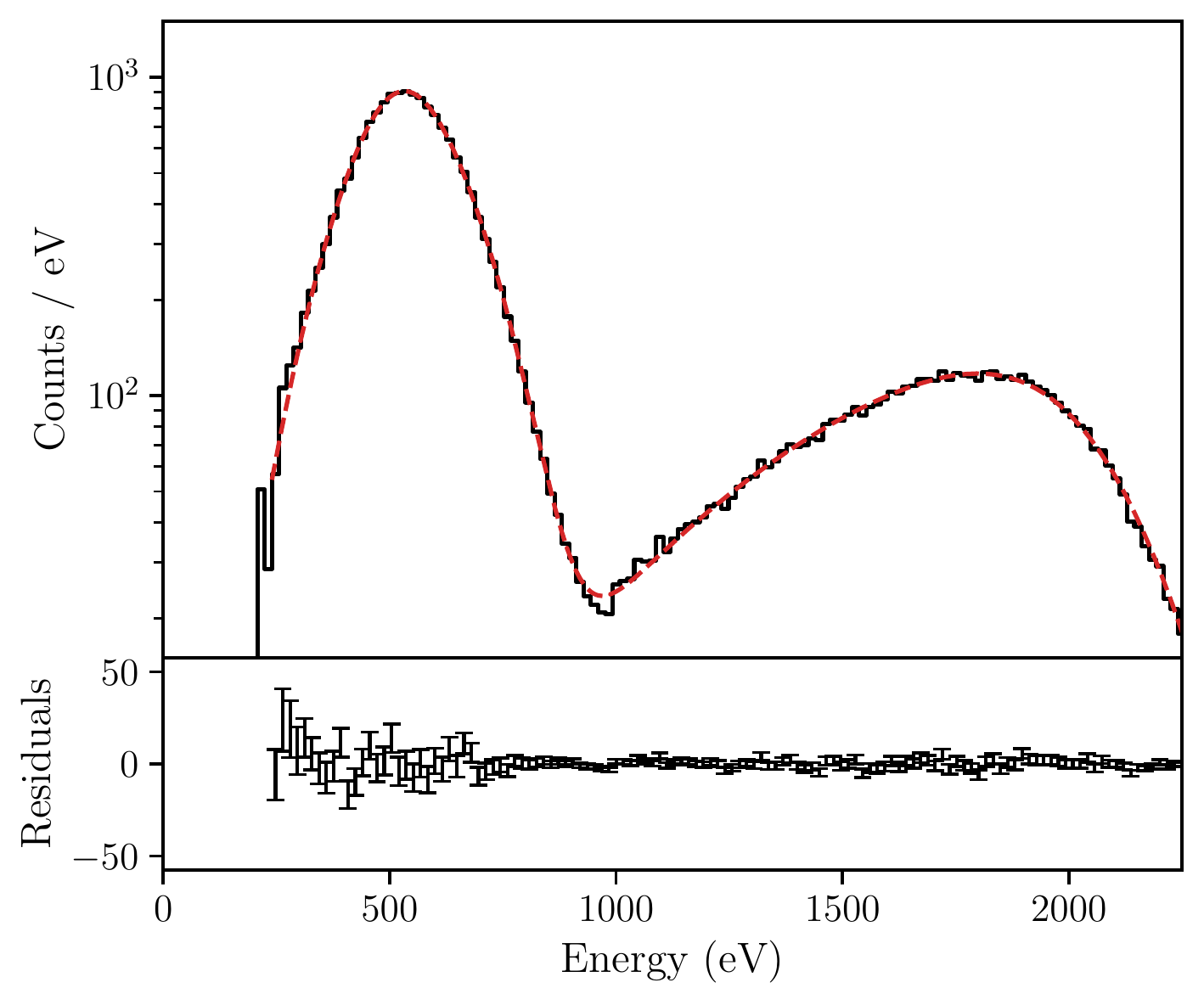}%
    }\quad
    \subfloat[PC]{%
      \includegraphics[width=0.48\textwidth]{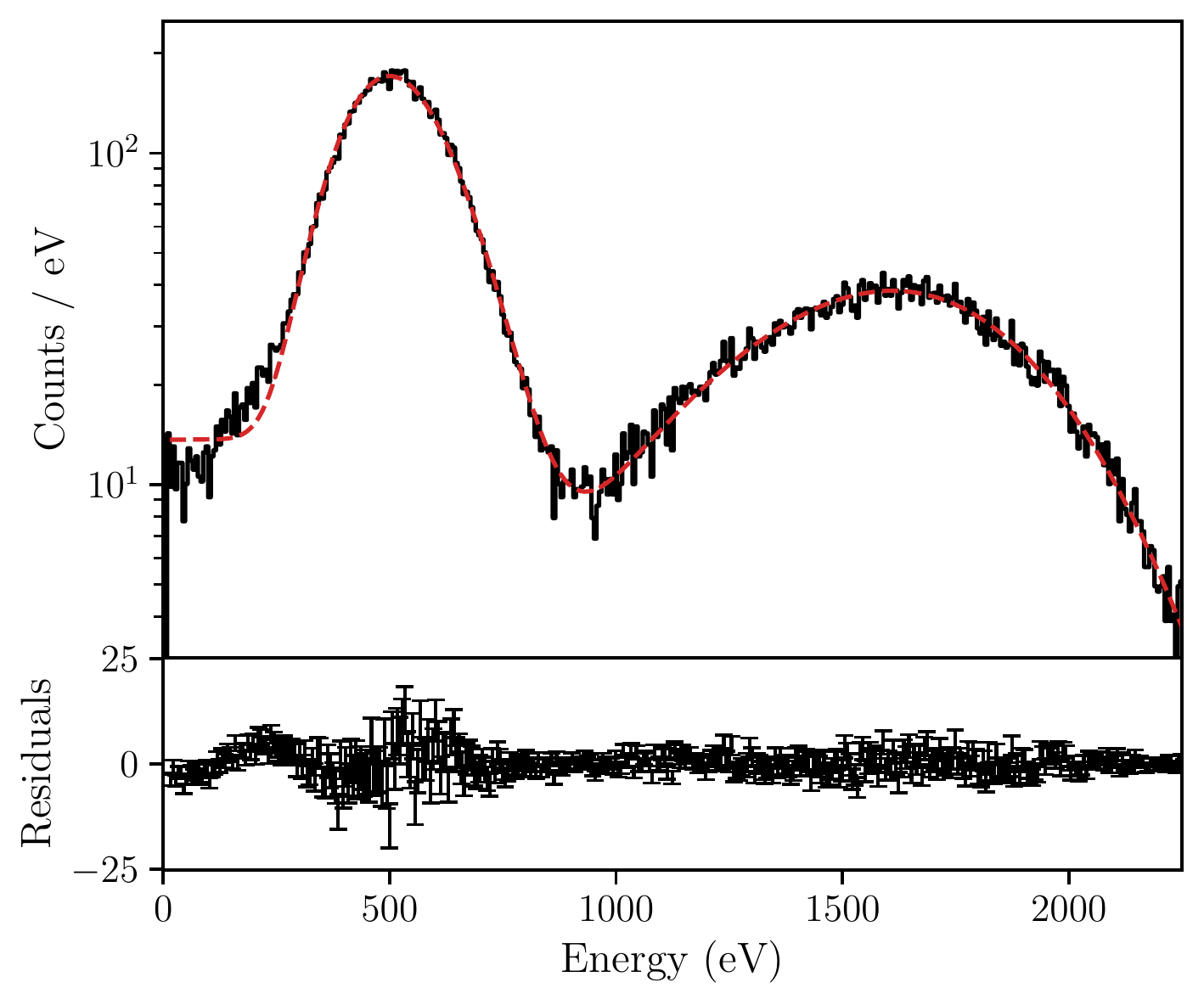}
    }
    \vspace{0.25 cm}
    \caption{H2RG  and PC spectra from the O \ka QE measurements.  The fitted models are shown in red with residuals plotted below. Similar to the Al \ka spectrum, the low-energy peaks represent the O \ka counts while the higher-energy peaks represent continuum emission (blended with Mg \ka counts).\vspace{0.4 cm}}
    \label{fig:spec_O}
\end{figure}

The Al \ka and O \ka spectra contain continuum emission from the Manson source in addition to the characteristic X-ray peaks. This continuum emission must be separated out to obtain accurate QE measurements. 
For this reason, we chose to model and fit these spectra, allowing us to use only the model components corresponding to the characteristic emission in our QE analysis. 
The models, described below, were fit using a least-squares algorithm.
The total number of counts was found by integrating the area under the fitted curves of the model components corresponding to characteristic X-ray peaks.
Figs.~\ref{fig:spec_Al}~\&~\ref{fig:spec_O} shows the Al \ka and O \ka spectra along with the fitted models and residuals.

We removed background counts from the H2RG spectra before fitting by subtracting the binned counts of the dark spectra from those of the Al \ka and O \ka spectra. 
The Al \ka and O \ka H2RG peaks were modeled as the sum of two Gaussian components.
The higher-energy continuum peaks (blended with an additional Mg \ka peak in the O \ka measurement) were modeled as skew-normal curves. 

The PC spectra were modeled using the distributions described in Ref.~\citenum{Auerhammer98}. 
The characteristic peaks were modeled as Prescott functions \cite{Prescott63}:
\begin{equation}
    P(x) = \frac{P_0\;x_0^{1/4} }{x^{3/4}\sqrt{4\pi Q}} \exp{\left[-\left(\sqrt{x}-\sqrt{x_0}\right)^2/Q\right]},
    \label{eq:prescott}
\end{equation}
where $x_0$ is a parameter which defines the location of the peak, $Q$ defines its width, and $P_0$ defines its amplitude. 
The lower-energy PC spectra also feature a shelf below the characteristic peaks. 
This shelf is the result of events with partial charge collection, where a fraction of the charge recombines before amplification and collection. 
The low-energy shelves were modeled as the following: 
\begin{equation}
    S(x) = S_0 \; \left(1-\textrm{erf}\left[\frac{x-x_0}{2\sqrt{2} \, x_0 \, Q}\right]\right)
    \label{eq:shelf}
\end{equation}
where $x_0$ and $Q$ are parameters from the Prescott function and $S_0$ is a parameter defining the amplitude of the shelf. 
As in the H2RG spectra, the higher-energy continuum peaks are modeled as skew-normal curves.
We observed that the PC background was fairly constant with energy (as seen in Fig.~\ref{fig:spec_Mn}b) and modeled it as a constant.

\subsection{QE Calculation}

The QE is given by the ratio of the flux measured by the H2RG to the flux incident on the detector. 
We infer the incident flux on the H2RG by scaling the flux incident on the PC (given by $F_\textrm{\tiny{PC}}/\textrm{QE}_\textrm{\tiny{PC}}$) by a factor ($k$) accounting for the difference in beam intensity at the locations of the detectors.
The value of $k$, which gives the ratio of the flux incident on the H2RG to that incident on the PC, is determined using either the Manson beam calibration measurements (described in \S\ref{sec:manson}) or the distances of the detectors from the Fe-55 source.  
These factors are combined to obtain the following expression for the H2RG QE:
\begin{equation}
    \text{QE}_\text{\tiny H2RG} = \text{QE}_\text{\tiny PC} \frac{F_\text{\tiny H2RG}}{k \; F_\text{\tiny PC}}.
    \label{eq:qe_calc}
\end{equation}

\subsection{Error Analysis}

We computed the error on our QE measurements using a Monte Carlo analysis, in which we repeatedly sampled input variables from their respective distributions in order to sample the QE error distribution.
These input variables include the observed count rates, PC dimensions and gas properties, and incident flux and PC window calibration measurements.
We used $10^4$ simulations to obtain a large enough sample to provide accurate error estimates.
We used a Shapiro-Wilk test \cite{Shapiro65} to test for the normality of the resulting distributions and found that the errors for all three of the QE measurements were consistent with a Gaussian distribution with P-values greater than 0.05.  
All uncertainties provided in this paper are the $1\sigma$ values taken from these distributions.

\begin{table}[bt]
    \caption{Values used for the calculation of the H2RG QE. Flux is given in units of counts s$^{-1}$ cm$^{-2}$.}
    \begin{center}
    \begin{tabular}{lcccccc}
    \toprule
    \textbf{Emission Line}  & \textbf{H2RG Flux} & \textbf{PC Flux} & \textbf{PC QE} & \textbf{Incident Flux Ratio} & \textbf{Measured QE}\\
    \midrule
    Mn K$\upalpha$/K$\upbeta$ &  $18.41 \pm 0.06$  & $13.41 \pm 0.07$ & $69.3 \pm 0.7\%$ & $1.0018 \pm 0.0003$ & $95.0 \pm 1.1\%$\\
    Al \ka      & $51.46 \pm 0.11$& $39.49 \pm 0.18$  & $74.9 \pm 1.1\%$ & $0.991 \pm 0.010$ & $98.5 \pm 1.8\%$ \\
    O \ka      &  $42.58 \pm 0.14$ & $19.25 \pm 0.10$ & $38.5 \pm 1.2\%$ & $1.000 \pm 0.002$ & $85.0 \pm 2.8\%$ \\
    \bottomrule
    \end{tabular}
    \label{tab:QE_calc}
    \end{center}
\end{table}

\break
\subsection{QE Results}
\label{sec:results}

\begin{figure}[bt]
    \centering
    \includegraphics[width=0.65\textwidth]{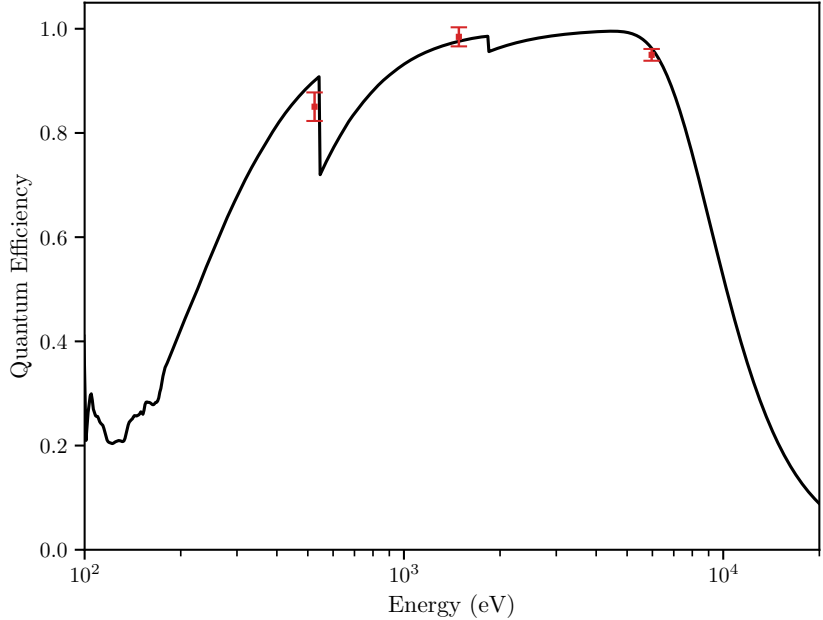}
    \caption{Measured QE values plotted on top of the slab-absorption H2RG QE model. The Fe-55 measurement is plotted at the weighted mean of the Mn \ka and \kb energies. \vspace{0.4 cm}}
    \label{fig:qe_res}
\end{figure}

Table~\ref{tab:QE_calc} shows the QE results at each of the measured energies, along with the fluxes, PC QE, and incident flux ratios used in the calculations. 
We measured the effective QE to be $95.0\pm1.1\%$ at the Mn K$\upalpha$/K$\upbeta$ energies, $98.5\pm1.8\%$ at the Al \ka energy, and $85.0\pm2.8\%$ at the O \ka energy. 
The Mn K$\upalpha$/K$\upbeta$ and Al \ka measurements are both in good agreement with the previous QE measurements made on H1RG HCDs \cite{Bongiorno15, Prieskorn14}.
Many of these H1RGs have Al blocking filters, but have the same absorber layer thickness ($100\;\upmu$m); therefore, they should have similar QEs at moderate energies.
All of these measurements represent high QE, comparable to other state-of-the-art X-ray detectors and meet the QE requirements for use in Lynx ($\geq85\%$\cite{Falcone19}). 

Table~\ref{tab:QE_res} shows the measured QE values compared with the values predicted by the model (previously described in \S\ref{sec:model}). 
The measurements are also shown plotted over the QE model in Fig.~\ref{fig:qe_res}.
The Mn K$\upalpha$/K$\upbeta$ value presented in the table is the expected effective QE, computed by taking the weighted average of the modeled QE at the Mn \ka and \kb energies (using the expected Mn K$\upalpha$/K$\upbeta$ ratio specific to our source). 
At each of the measured energies, the measurements are in reasonable agreement with the model values, indicating that the 1d slab-absorption model performs well at low and moderate energies.

We plan to measure the QE at the the carbon \ka line (0.28 keV), in order to further investigate the performance of this detector at even lower energies and to ensure that the QE is not degraded by effects such as charge diffusion or variations in the oxide layer thickness.
We also plan to use a similar experimental setup to measure the QE of the Speedster\cite{Griffith16, Burrows19} and small-pixel \cite{Hull19} HCDs, detectors with the ability to provide the high frame rate and spatial resolution required for Lynx and other future missions.

\begin{table}[bt]
    \caption{Comparison of H2RG QE measurements with the expected model values.}
    \begin{center}
    \begin{tabular}{lccc}
    \toprule
    \textbf{Emission Line} & \textbf{Energy (keV)} & \textbf{Measured QE} & \textbf{Model QE}\\
    \midrule
    Mn K$\upalpha$/K$\upbeta$ & 5.90 / 6.49 & $95.0 \pm 1.1 \%$ & 96.2\% \\
    Al \ka     & 1.49        & $98.5 \pm 1.8 \%$ & 97.6\% \\
    O \ka      & 0.52        & $85.0 \pm 2.8 \%$ & 90.0\% \\
    \bottomrule
    \end{tabular}
    \label{tab:QE_res}
    \end{center}
\end{table}

\break
\section{CONCLUSION}
\label{sec:conclusion}

We have measured the quantum efficiency of a Teledyne Imaging Sensors H2RG HCD at the Mn K$\upalpha$/K$\upbeta$, Al K$\upalpha$, and O \ka energies using the X-ray beamline at Penn State University.
The measured QE values are in good agreement with our model, which is based on the absorption of different detector layers. 
Our results show that hybrid CMOS detectors are capable of achieving high QE across the soft X-ray band, meeting a crucial requirement of proposed future X-ray observatories.
We plan to extend our measurements to the C \ka line to further test the low-energy capabilities of this detector.
We also plan to make QE measurements on next-generation Speedster and small-pixel detectors, which incorporate some of the latest design improvements that further increase the capabilities of X-ray hybrid CMOS detectors.


\acknowledgments 
 
This work was supported by a NASA Space Technology Graduate Research Opportunity, as well as NASA grant 80NSSC20K0778.  We would also like to acknowledge useful discussions and advice from Yibin Bai at Teledyne Imaging Sensors and Ben Zeiger at Luxel.

\break

\bibliography{report} 
\bibliographystyle{spiebib} 

\end{document}